\documentclass[conference]{IEEEtran}
\IEEEoverridecommandlockouts
\usepackage{cite}
\usepackage{amsmath,amssymb,amsfonts}
\usepackage{algorithmic}
\usepackage{graphicx}
\usepackage{textcomp}
\usepackage{xcolor}
\usepackage[hyphens]{url} 
\def\BibTeX{{\rm B\kern-.05em{\sc i\kern-.025em b}\kern-.08em
    T\kern-.1667em\lower.7ex\hbox{E}\kern-.125emX}}
\begin{document}

\title{M2LADS: A System for Generating MultiModal Learning Analytics Dashboards in Open Education\\
}

\author{\IEEEauthorblockN{
Álvaro Becerra\IEEEauthorrefmark{1},
Roberto Daza\IEEEauthorrefmark{1},
Ruth Cobos\IEEEauthorrefmark{1},
Aythami Morales\IEEEauthorrefmark{1},
Mutlu Cukurova\IEEEauthorrefmark{2} 
and
Julian Fierrez\IEEEauthorrefmark{1}}

\IEEEauthorblockA{\IEEEauthorrefmark{1}School of Engineering, Universidad Autonoma de Madrid, Spain\\
\{alvaro.becerra, roberto.daza,
ruth.cobos, aythami.morales, julian.fierrez\}@uam.es}

\IEEEauthorblockA{\IEEEauthorrefmark{2}University College London, London, UK\\
\{m.cukurova\}@ucl.ac.uk}

}

\maketitle

\begin{abstract}
In this article, we present a Web-based System called M2LADS, which supports the integration and visualization of multimodal data recorded in learning sessions in a MOOC in the form of Web-based Dashboards. Based on the edBB platform, the multimodal data gathered contains biometric and behavioral signals including electroencephalogram data to measure learners' cognitive attention, heart rate for affective measures, visual attention from the video recordings. Additionally, learners' static background data and their learning performance measures are tracked using LOGCE and MOOC tracking logs respectively, and both are included in the Web-based System. M2LADS provides opportunities to capture learners' holistic experience during their interactions with the MOOC, which can in turn be used to improve their learning outcomes through feedback visualizations and interventions, as well as to enhance learning analytics models and improve the open content of the MOOC.
\end{abstract}


\section{Introduction}
Open education encompasses a wide range of practices, including the use of open educational resources (OER)~\cite{chan2020enhancing}, open access journals, open textbooks, and open courseware. It also involves the use of open technologies and platforms, such as MOOCs (Massive Open Online Courses), which provide free and open access to education for anyone with an internet connection.

MOOCs are a valuable source of educational content and are sometimes endorsed and recognized by official institutions~\cite{ma2019investigating}. But, how do these MOOC learners behave? How do they interact with the courses? What is their learning context like? Are they focused when learning?

The research area that could help us answer these questions is Multimodal Learning Analytics (MMLA), which is a subfield of Learning Analytics~\cite{lang2017handbook, martinez2020achievements, romeroeducational} that deals with collecting and integrating data from different sources, providing a panoramic understanding of the learning processes and the various dimensions related to learning~\cite{giannakos2022multimodal}.

Several attempts have been made to integrate multimodal data collected from various sources in educational contexts, as discussed in the following section. However, due to the diversity and volume of biological and behavioral signals that we aim to monitor in MOOCs, and considering the unique characteristics of the platform where we intend to track learner data, we have opted to create a web-based system. This system aims to facilitate the accurate integration and visualization of multimodal data during learning sessions in MOOCs through web-based dashboards.

The system is called M2LADS, which is an acronym for ''System for Generating Multimodal Learning Analytics Dashboards''. Using these dashboards, MOOC instructors, as users of this approach, can perform in-depth analysis to gain a better understanding of when MOOC learners are focused, what course content captures their attention, and more.. Furthermore, this work is based on several learning analytics tools~\cite{cobos2016open, cobos2020proposal, cobos2021improving,pascual2022proposal} to address related challenges.

The structure of this article is as follows: in the next section, we present related works. In the following section, we provide a detailed description of the proposed system. In section~\ref{sec:Case Study}, we present the current use of the system in a case study with learners of a MOOC on edX\footnote{\url{https://www.edx.org/}}. Finally, the article concludes with conclusions and future work.

\section{Related Work}

In recent years, within the field of Learning Analytics, dashboards have been developed to assist both learners and instructors~\cite{karademir2022designing, verbert2014learning}. Their goal is to collect personal information about learners, including their progress in a course, among other data, and display it through graphical representations as Learning Analytics Dashboards (LADs)~\cite{matcha2019systematic}. In the case of learners, these dashboards can help them make better decisions regarding their learning. In the case of instructors, the goal of these dashboards is to help them optimize and improve their teaching.

As we can see, information such as visual attention, cognitive load, or stress is of great value for dashboards since they allow for an improvement in the learning process. Not only has this multimodal information demonstrated great value for the educational branch, but also biometric information allows for the identification of possible suspicious and fraudulent events, as well as uncanny behaviors that can be observed by instructors to guarantee safe education or certification~\cite{hernandez2019edbb,baro2018integration,morales2016kboc}.

Similar analyses have also been developed in other studies, such as in~\cite{sharma2020eye}, where the readings from a MOOC are analyzed to visualize learner areas of interest through fixations and heat maps, and subsequently predict what has been learned. Another example can be found in~\cite{ha2015tracking}, where instead of a MOOC, a set of LA dashboards is used in which each of the graphs is defined as areas of interest, and an eye-tracker is used to obtain fixations in each of them.

Capturing multimodal data is characterized by the use of a large number of sensors, devices, or signals \cite{2018_INFFUS_MCSreview1_Fierrez}. Several studies~\cite{giannakos2019multimodal,spikol2018supervised} have shown that better prediction results for learner success are obtained by using multiple data acquisition devices. An approach where multimodal data from multiple devices is monitored, analyzed, and visualized is the research presented by Acevedo and colleagues in the area of Intelligent Tutoring~\cite{Azevedo2022}.

We have found some related and interesting research works in the area of multimodal discourse analysis of collaborative knowledge construction~\cite{ouyang2023artificial} and multimodal analysis of nonverbal aspects of collaboration~\cite{cukurova2020modelling}. As these research areas are concerned with understanding how different modes of communication, such as language, gesture, and visual images, work together to create meaning in collaborative settings, capturing multimodal data from multiple devices has been demonstrated to be very useful.

However, the development of dashboards often focuses only on a few devices, such as the eye-tracker, and does not address any of the themes proposed in~\cite{matcha2019systematic} that are used in Learning Analytics Dashboards (LADs), such as aiding in the detection of learning problems and difficulties or helping learners in the development of a learning plan.

This limitation, where there are not enough approaches to help visualize and analyze multimodal data from different devices in online courses, has motivated the creation of the proposal that we present in this article.

\begin{figure}[t]
 \centering
  \includegraphics[width=\linewidth]{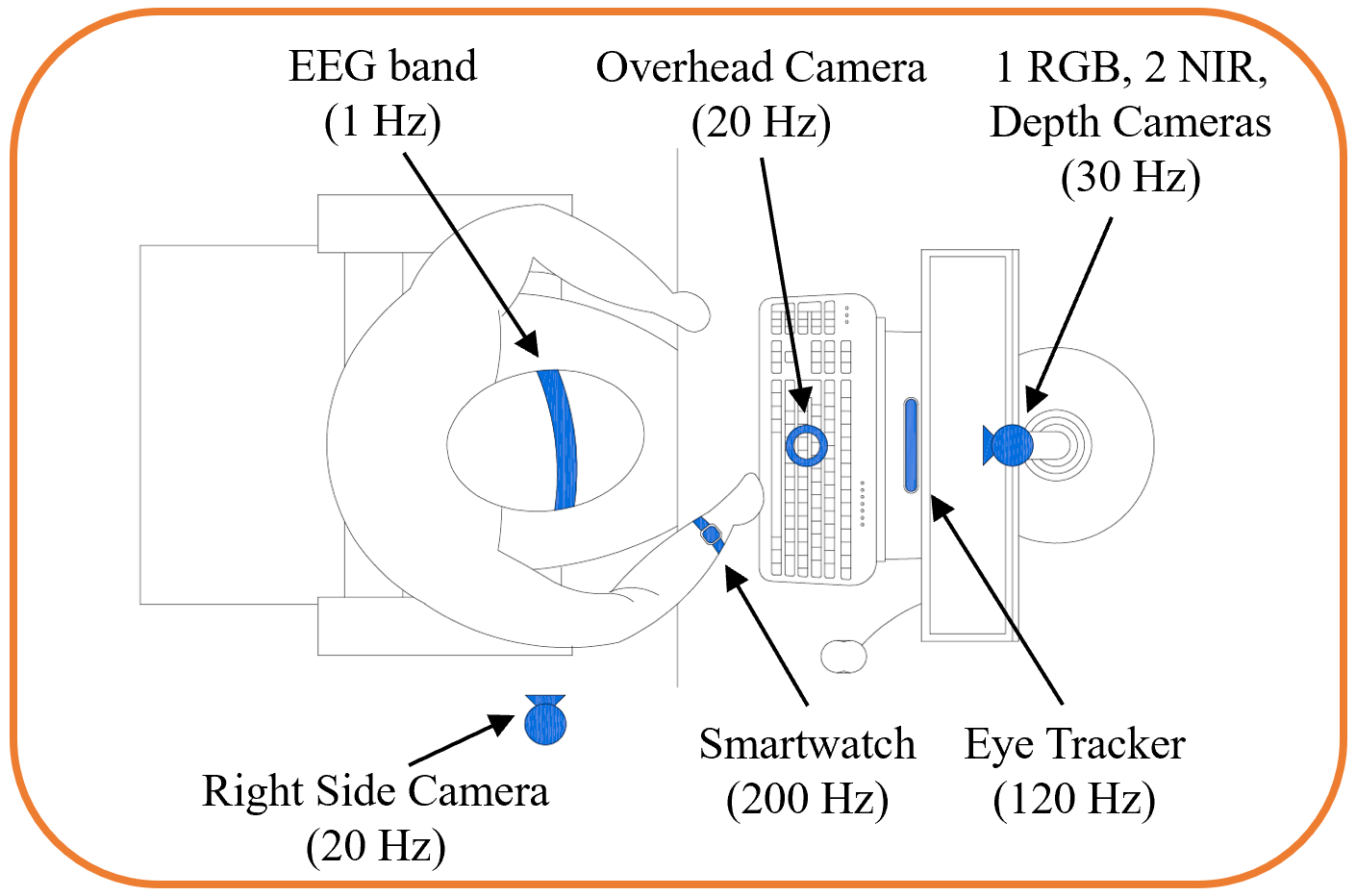} 
  \caption{The acquisition setup shows how it’s been setup for student’s monitoring in remote education, using the edBB platform~\cite{hernandez2019edbb}. For each sensor, the Sampling Rate used is shown.}
   \label{Setup_edBB}
\end{figure}

\section{The approach: M2LADS}
We propose a web-based system called M2LADS (an acronym for System for generating Multimodal Learning Analytics Dashboards). This system supports the generation of web-based dashboards. Each dashboard visualizes all the multimodal data monitored in a learning session (LS) in a MOOC. We have developed the system for any edX MOOCs; however, we have tested it on a specific course (see Section~\ref{sec:Case Study} for more details). These dashboards can be accessed and viewed by the MOOC instructors due to they are the potential users of this approach.

The system is composed of three modules following a Model-View-Controller (MVC) approach. For simplicity, the multimodal data recorded in the LS and processed by the system is referred to in this approach as Activity Data (AD). Details of this multimodal data are found in Section~\ref{sec:MultiModal Data Description}.

The Modules of M2LADS System are:

\begin{itemize}
 
\item Activity Data Processing Module (Controller): see  Section~\ref{sec:AD Processing Module}.

\item Activity Data Store Module (Model): see Section~\ref{sec:AD Management Module}.

\item Activity Data Visualization Module (View): see Section~\ref{sec:AD Visualization Module}.
\end{itemize}

\subsection{MultiModal Data Description} \label{sec:MultiModal Data Description}

\subsubsection {edBB Data}
The EdBB platform was used to monitor learners during e-learning activities~\cite{daza2023edbb,hernandez2019edbb}. This multimodal acquisition framework captures biometric and behavioral information and was designed for remote education. It uses a group of software to communicate and synchronize different sensors and can adapt the acquisition setup to the monitoring circumstances, using both advanced sensors (smartwatch, eye tracker, etc.) and basic ones (webcam, context data, etc.). In our work, we used the following acquisition setup and sources of information/sensors (see Fig.~\ref{Setup_edBB}):
\begin{itemize}
\item Video: Video data from $3$ different positions: Overhead, front and side cameras, using $2$ simple webcams and $1$ Intel RealSense that includes $1$ RGB and $2$ NIR cameras; that also calculates the depth images using the NIR cameras. Lastly, the screen monitoring video is also captured.
\item Electroencephalogram (EEG) data: Using a NeuroSky EEG band that obtains $5$ signals: $\delta$~($<4$Hz), $\theta$~($4$-$8$ Hz), $\alpha$~($8$-$13$ Hz), $\beta$~($13$-$30$ Hz), and $\gamma$~($>30$ Hz) and through the pre-processing of these EEG channels, attention, meditation, and the moment in which the blinks occur, are also obtained~\cite{daza2022alebk,daza2023matt,daza2020mebal}. 
\item Heart rate: To capture this in real time we use a Huawei Watch $2$ pulsometer feature~\cite{hernandez2020heart}.
\item Visual attention: A Tobii Pro Fusion was used and it contains two eye tracking cameras that estimate the following data: Gaze origin and point, pupil diameter, data quality, etc.; allowing us to obtain visual attention.
\item Metadata: We collect it from sessions and PCs like IP and MAC addresses, OS, apps and web history.
\end{itemize}

\begin{figure*}[t]
 \centering
  \includegraphics[width=\linewidth]{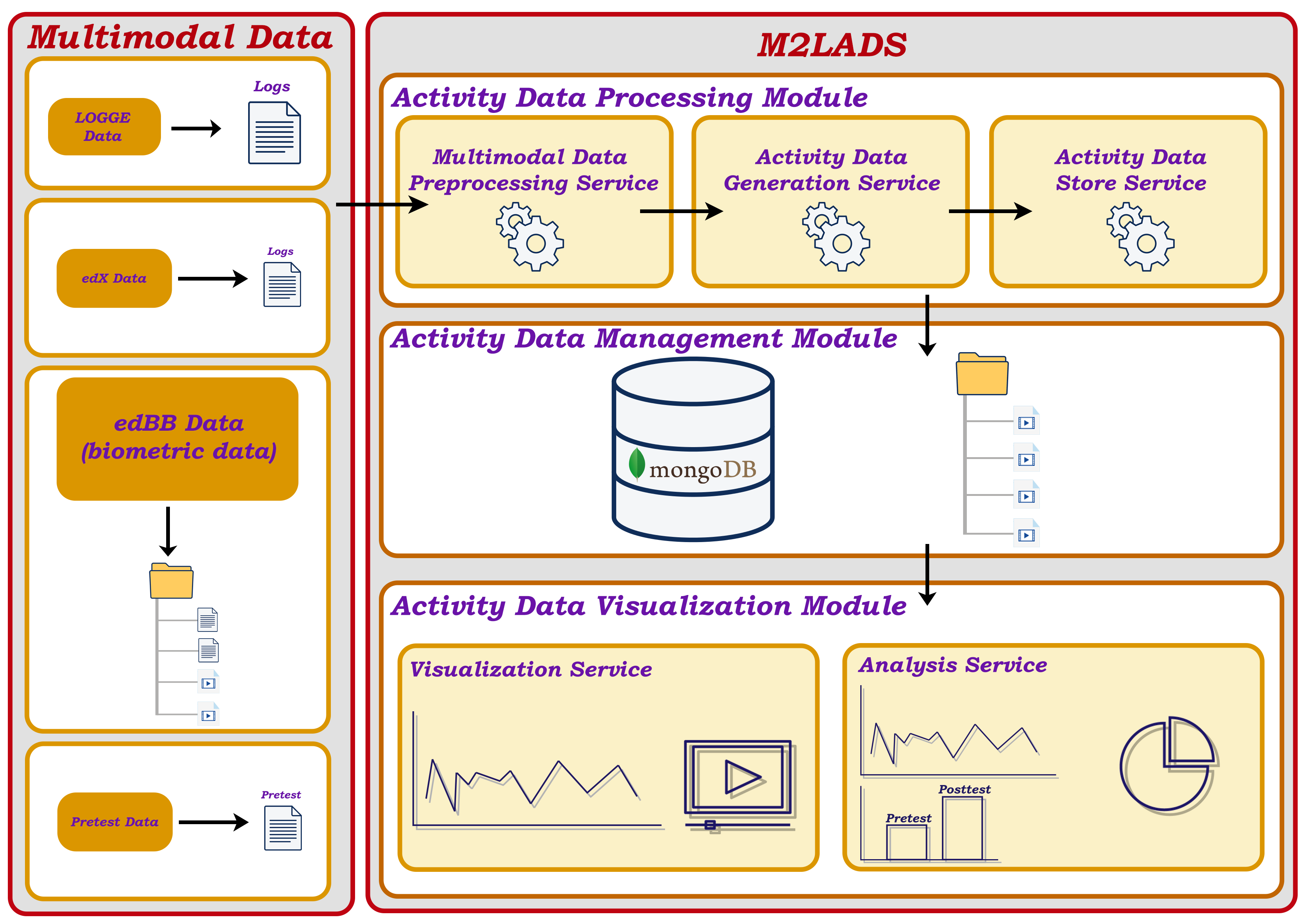} 
  \caption{M2LADS Architecture/Modules}
   \label{arquitectura_M2LAD}
\end{figure*}

\subsubsection {MOOC Data}
While a learner is learning in a MOOC on edX, log data is generated. This log data contains detailed information about where the learner is navigating, which videos he/she visits, which assignments he/she completes, and so on.

All this information is correctly contextualized with the learner's data and the time stamp of each action. The log data of a LS is stored in a JSON file.

\subsubsection {Additional Data}
Despite the large amount of information provided by edX log data, a complementary log recording tool called LOGGE has been developed in the context of this approach.

The objective of this tool is to store additional information related to the monitored learner (e.g., sex, hand used with the mouse, heart problems, marks, etc.) and enrich the edX log traces with information related to the activities that the learner carries out during the LS. This information (log data) is stored as a CSV file.

Furthermore, to assess learner learning, a pretest related to the MOOC LS has been designed. This pretest is answered by learners just prior to the start of the LS.

\subsection{Activity Data Processing Module} \label{sec:AD Processing Module}

\begin{figure}[t!]
\centering
    \includegraphics[width=\linewidth]{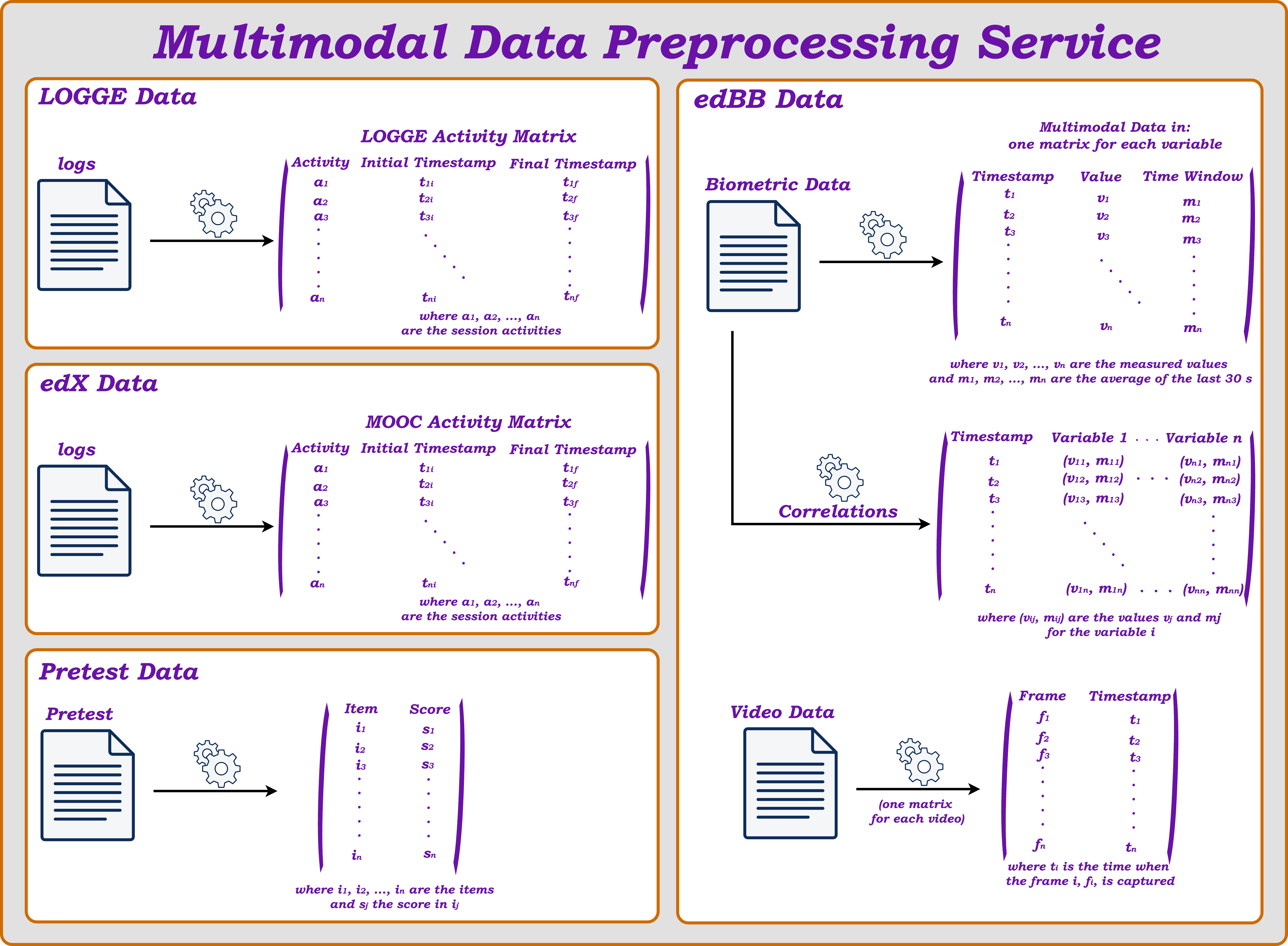}
\caption{Multimodal Data Preprocessing Service}
\label{MMDPS}

\end{figure}

This module allows for the extraction, cleaning, selection, and preprocessing of the multimodal data recorded during the LS in the MOOC to extract the learner's Activity Data. The module is composed of the following three services:

It is designed to extract and process data from these four sources (see Fig.~\ref{MMDPS}):
\begin{itemize}
\item LOGGE Data: The log file generated by the LOGGE tool, which contains information related to the activities or events that the learner has carried out and their timestamps, is preprocessed into an activities matrix (LOGGE Activity Matrix) with three columns: the activity identifier, the initial timestamp, and the final timestamp of each activity.
\item edX Data: The edX log file is cleaned and preprocessed into another activities matrix (MOOC Activity Matrix) with the same structure as the previous one.
\item Pretest Data: The learner's answers to the pretest items are checked against the correct answers, and a matrix with two columns is generated: item and score.
\item edBB Data: The biological and behavioral signals provided by biometric devices are preprocessed as follows: for each variable, such as attention (from EEG), heart rate (from smartwatch), and pupil diameter (from eye-tracker), a matrix is generated with these columns: timestamp, variable value, and time window (the average of the most recent data within the last 30 seconds). In all of these matrices, the time unit is standardized, and the start and end times are synchronized so that the data for all variables start and end simultaneously. In addition, correlations are calculated. For each video, frames, and the timestamps at which those frames were taken are extracted, and a matrix per video is generated with this information in two columns.

\end{itemize}

\subsubsection{Activity Data Generation Service} 
\begin{figure}[t!]
\centering
    \includegraphics[width=\linewidth]{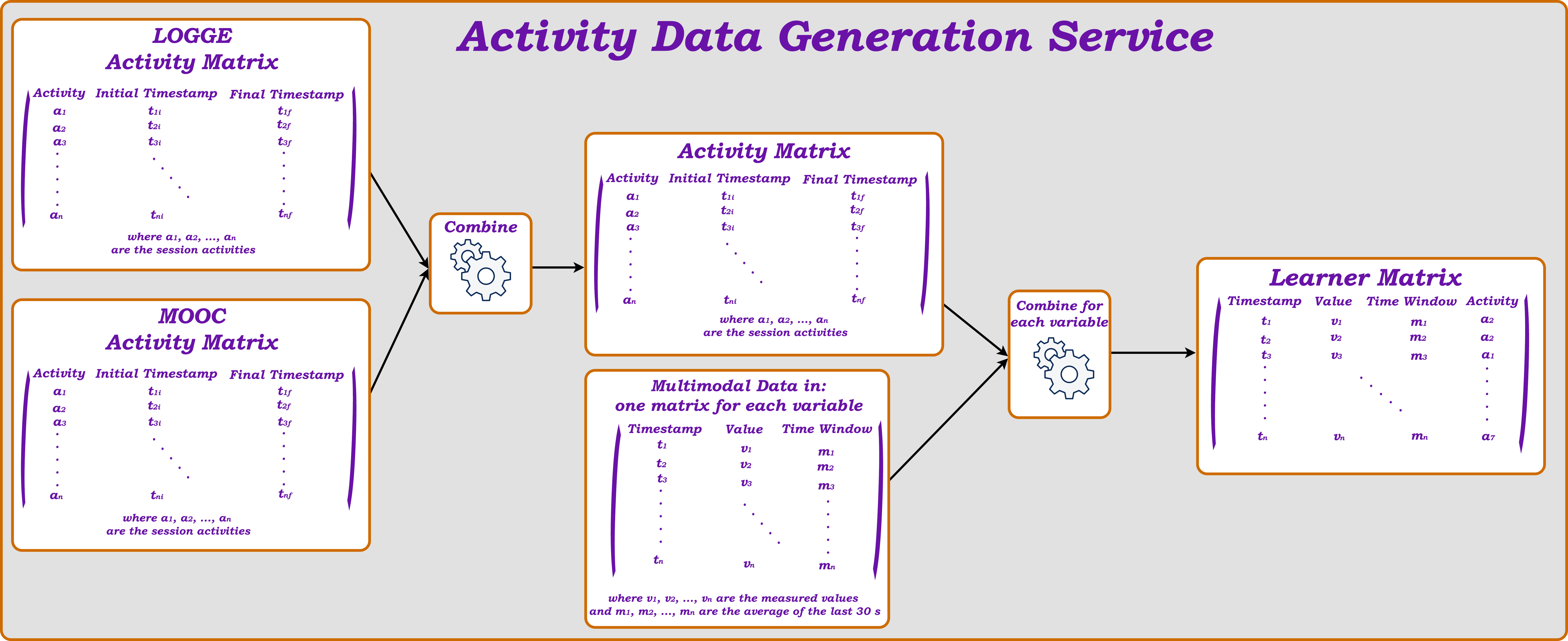}
\caption{Multimodal Data Preprocessing Service}
\label{matrices}

\end{figure}

This service crosses the data from the previous learner data organized in four matrices to generate the final Learner Matrix (LM) following this process (see Fig.~\ref{matrices}):

Firstly, the LOGGE Activity Matrix and the MOOC Activity Matrix are combined into an intermediate matrix (Activity Matrix). If an activity is present in only one of the initial matrices, its data is included in the intermediate matrix. However, if an activity is present in both initial matrices, the data from the MOOC matrix is given priority and included in the intermediate matrix. Finally, each of the matrices with the data for each of the variables (attention, heart rate, etc.) is combined with the Activity Matrix. This way, the information on what activity the learner was doing while each biometric value was being recorded is obtained and included in the final Learner Matrix (LM), which contains four columns: timestamp, variable value, time window, and the activity identifier.
\subsubsection{Activity Data Store Service}

This service stores the LM as data collections in a MongoDB database, and all the recorded videos during the LS are organized as audiovisual files in a set of directories.

\subsection{Activity Data Management Module} \label{sec:AD Management Module}
This module provides connectivity with MongoDB and the directories with the audiovisual files.

\subsection{Activity Data Visualization Module}\label{sec:AD Visualization Module}

In this module, the system creates a visualization per learner, i.e., a dashboard, that reflects learner activity data during the LS. To achieve this, it generates and organizes visual components (graphs) using the Dash framework\footnote{\url{https://plotly.com/dash/}}, which is based on Flask and React.js.

\begin{figure}[t]
 \centering
  \includegraphics[width=\linewidth]{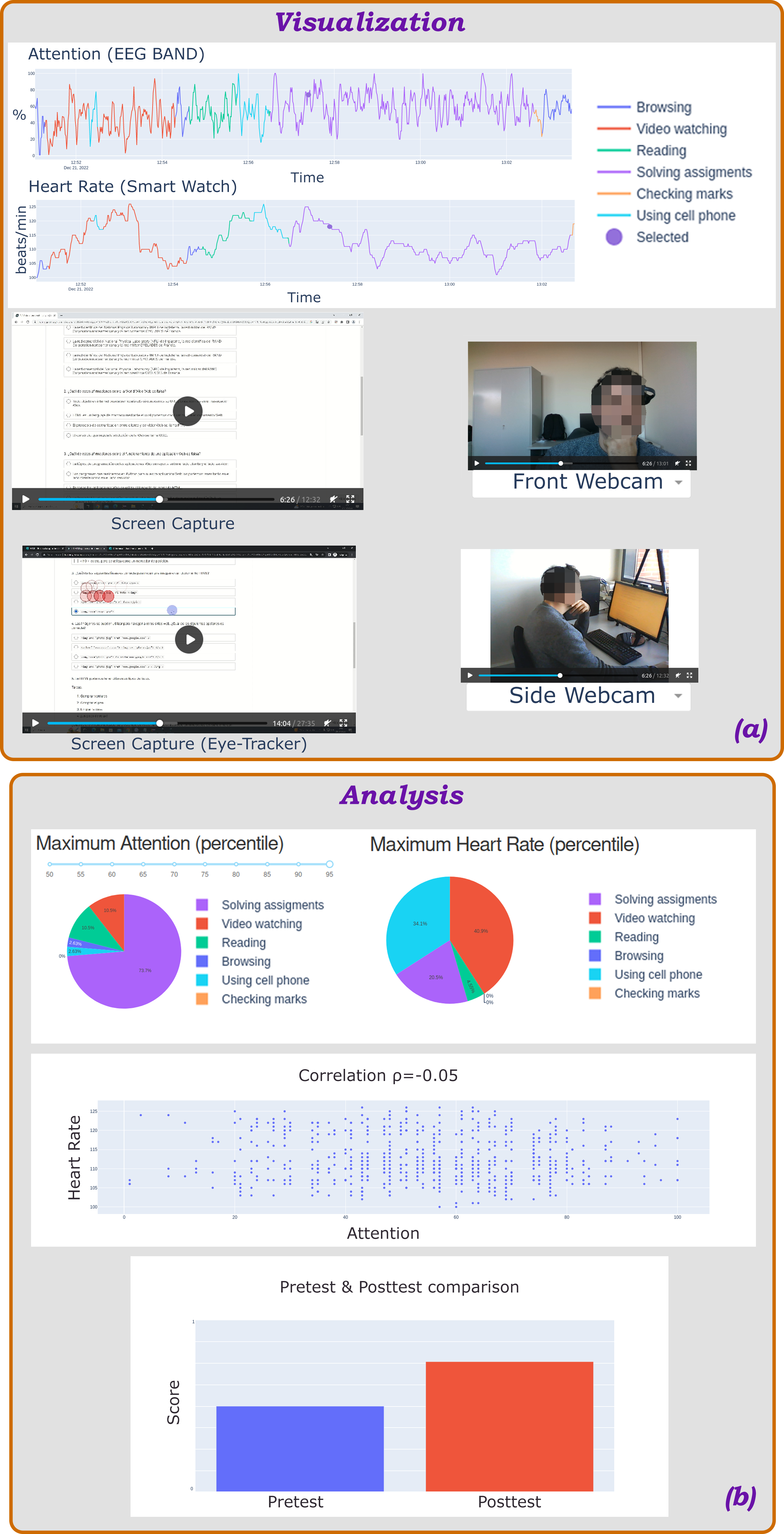} 
\caption{Screenshot of some visualization by M2LADS}
\label{FRONT_END_M2LAD}
\end{figure}

\subsubsection{Visualization Service}  

As shown in Fig. \ref{FRONT_END_M2LAD} (a), this service composes the dashboard with several graphs that display the learner's attention, meditation, heart rate, and neural waves over the course of the LS. It also includes several videos, such as the screen captured during the LS, different webcam videos, and learner fixation areas video from the eye-tracker. All these elements in the dashboard are synchronized.

\subsubsection{Analysis Service} Fig. \ref{FRONT_END_M2LAD} (b) illustrates that this service adds graphs to the dashboard, which display an analysis of the previous data, correlations among the values of the biometric signals, and comparative values of learner performance (pretest score vs posttest score).

\section{Case Study} \label{sec:Case Study}
The presented approach was tested in the MOOC entitled "Introduction to Development of Web Applications"  (WebApp), which is available on the edX MOOC platform and offered by the Universidad Autonoma de Madrid (UAM). UAM joined the edX Consortium in 2014, and since then, it has been offering a wide variety of courses. We selected the WebApp MOOC because we needed its content for the learning scenario of our research study.

The course teaches learners
to develop Web Applications. Learners who acquire web application development skills through this MOOC they will have a valuable and in-demand skillset that can open up many opportunities in their career. The ability to develop web applications is critical for computer science engineers who work with technology and software, and it is an increasingly important skill in today's job market.

The course is divided into five units and covers these essential essential technologies and programming languages:  HTML, CSS, Python, JSON, JavaScript, and Ajax. Each unit consists of several subunits, and each subunit contains open multimedia resources such as videos, HTML pages, PDF files, discussion forums, and graded assignments.

Currently, we are conducting a research study to measure the effect of cellphones on the attention levels of WebApp MOOC learners during learning sessions. To achieve this, we are monitoring 120 volunteers from School of Engineering at UAM who are required to attend our MMLA laboratory and interact with the same course subunit during a 30-minute LS. M2LADS will generate a web-based dashboard for each user, and each monitoring session will yield approximately 25 GB of information obtained from different channels such as biometric, behavioral, and metadata data. This study is endorsed by the university's ethics committee.

For this study, we have developed two initial research lines using the information provided by edBB and M2LADS:

\begin{itemize}
\item Technology development to detect the estimation of attention level based on video processing ~\cite{daza2022alebk,daza2023matt,daza2020mebal} and the analysis of cognitive load effects when students interact with mobile phones \cite{2020_CDS_HCIsmart_Acien} while in learning sessions.
\item Study based on the visual attention that the student holds while using a mobile phone, using automatic gaze targeting estimation technologies~\cite{chong2020detecting} obtained directly from a webcam. This allows the creation of visual attention maps that improve the information provided to M2LADS.
\end{itemize}

\begin{figure}
\includegraphics[width=\linewidth]{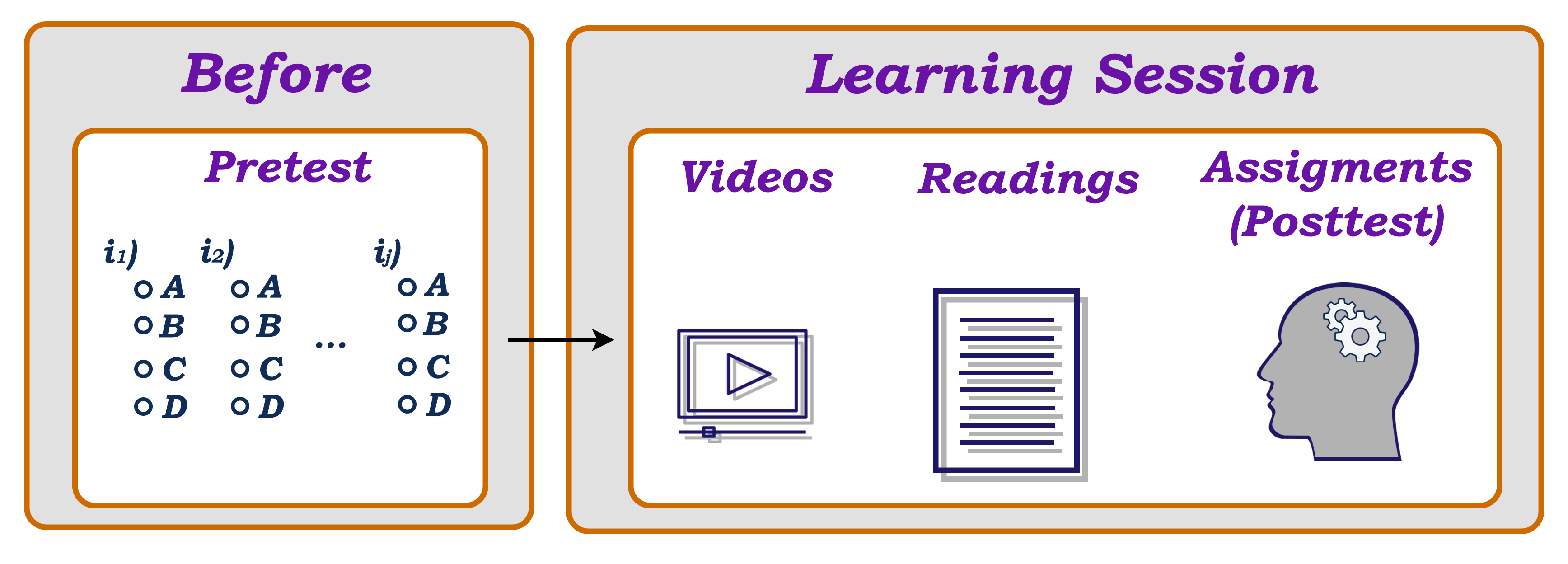}
\caption{Learning Session schema} \label{diagram_ls}
\end{figure}

As shown in Fig.~\ref{diagram_ls}, before starting with the collection of multimodal data from the LS, the learner takes a pretest with some items related with the LS topic to determine their initial level of knowledge. As seen in bottom part of Fig.~\ref{FRONT_END_M2LAD} (b),  the dashboard visualizes the learner's pretest performance compared to the performance obtained during the LS by answering assignments supported by the MOOC. These assignments are the items of the posttest. During the LS, the learner watches videos, reads materials, etc., as shown in Fig.~\ref{FRONT_END_M2LAD} (a).

\section{Conclusion and Future Work}\label{sec:Conclusion and Future Work}

In this article, we present a web-based system called M2LADS (an acronym for System for Generating Multimodal Learning Analytics Dashboards). This system supports the integration and visualization of multimodal data monitored in MOOC learning sessions in the form of web-based dashboards. These multimodal data are provided by the monitoring supported by the edBB platform and the data from the LOGGE tool and an edX MOOC.
Therefore, M2LADS’ main goal consists of providing valuable information from the sessions to the instructors, in order to make the learning system more personalized; ultimately, an improvement in education. 

edBB’ versatility allows the use of specific sensors like EEG, pulsometer, etc; to measure attention levels, heart rate, and other important signals. However, these unusual sensors can be replaced with deep learning technologies that work with images obtained from a basic webcam, that are capable of estimating attention levels~\cite{daza2022alebk,daza2023matt}, heart rate~\cite{hernandez2020heart}, eye blink detection~\cite{daza2020mebal}, user identification~\cite{daza2023edbb, morales2016kboc}, etc. Hence, with this secondary option, edBB can provide data for M2LADS in massive online education environments. 

Therefore, we can conclude that we have, at UAM, an MMLA laboratory ready with the integration of M2LADS, LOGGE tool, and edBB platform for new monitoring to exploit its utility in the e-learning context.

The initial results obtained from the proposed approach have been found to be useful by the team of instructors of a MOOC at edX (the online course being used in a research study conducted using this approach) in their tasks of improving the course content, particularly to detect which open resources in the form of videos would be advisable to improve.

Finally, future work will also explore new multimodal machine learning methods \cite{2023_SNCS_multiAI} capable of exploiting all the heterogeneous information sources \cite{2018_INFFUS_MCSreview2_Fierrez} originated in e-learning session can be properly combined with instructors and learners in human-in-the-loop AI-powered e-learning systems.

\section*{Acknowledgment}

Support by projects: HumanCAIC (TED2021-131787B-I00 MICINN), BIO-PROCTORING (GNOSS Program, Agreement Ministerio de Defensa-UAM-FUAM dated 29-03-2022),  IndiGo! (PID2019-105951RB-I00), SNOLA (RED2022-134284-T) and e-Madrid-CM (S2018/TCS-4307, a project which is co-funded by the European Structural Funds, FSE and FEDER). Roberto Daza is supported by a FPI fellowship from MINECO/FEDER.

\bibliographystyle{IEEEtran}
\bibliography{IEEEabrv,bibliography}

\begin{thebibliography}{10}
\providecommand{\url}[1]{#1}
\csname url@samestyle\endcsname
\providecommand{\newblock}{\relax}
\providecommand{\bibinfo}[2]{#2}
\providecommand{\BIBentrySTDinterwordspacing}{\spaceskip=0pt\relax}
\providecommand{\BIBentryALTinterwordstretchfactor}{4}
\providecommand{\BIBentryALTinterwordspacing}{\spaceskip=\fontdimen2\font plus
\BIBentryALTinterwordstretchfactor\fontdimen3\font minus
  \fontdimen4\font\relax}
\providecommand{\BIBforeignlanguage}[2]{{%
\expandafter\ifx\csname l@#1\endcsname\relax
\typeout{** WARNING: IEEEtran.bst: No hyphenation pattern has been}%
\typeout{** loaded for the language `#1'. Using the pattern for}%
\typeout{** the default language instead.}%
\else
\language=\csname l@#1\endcsname
\fi
#2}}
\providecommand{\BIBdecl}{\relax}
\BIBdecl

\bibitem{chan2020enhancing}
H.~C.~B. Chan, Y.-H. Ho, E.~Tovar, and S.~Reisman, ``Enhancing the learning of
  computing/it students with open educational resources,'' in \emph{Proc. IEEE
  COMPSAC}, 2020, pp. 113--122.

\bibitem{ma2019investigating}
L.~Ma and C.~S. Lee, ``{Investigating The Adoption of MOOC s: A
  Technology--User--Environment Perspective},'' \emph{Journal of Computer
  Assisted Learning}, vol.~35, no.~1, pp. 89--98, 2019.

\bibitem{lang2017handbook}
C.~Lang, G.~Siemens, A.~Wise, and D.~Gasevic, Eds., \emph{Handbook of Learning
  Analytics}.\hskip 1em plus 0.5em minus 0.4em\relax SOLAR, Society for
  Learning Analytics and Research New York, 2017.

\bibitem{martinez2020achievements}
A.~Mart{\'\i}nez~Mon{\'e}s, D.~Damoulis \emph{et~al.}, ``{Achievements and
  Challenges in Learning Analytics in Spain: The View of SNOLA},'' \emph{RIED.
  Revista Iberoamericana de Educaci{\'o}n a Distancia}, vol.~23, no.~2, p. 187,
  2020.

\bibitem{romeroeducational}
C.~Romero and S.~Ventura, ``{Educational Data Mining and Learning Analytics: An
  Updated Survey},'' \emph{Wiley Interdisciplinary Reviews: Data Mining and
  Knowledge Discovery}, 2020.

\bibitem{giannakos2022multimodal}
M.~Giannakos, D.~Spikol, D.~Di~Mitri, K.~Sharma, X.~Ochoa, and R.~Hammad, Eds.,
  \emph{The Multimodal Learning Analytics Handbook}.\hskip 1em plus 0.5em minus
  0.4em\relax Springer Nature, 2022.

\bibitem{cobos2016open}
R.~Cobos, S.~Gil, A.~Lareo, and F.~Vargas, ``{Open-DLAs: An Open Dashboard for
  Learning Analytics},'' in \emph{Proc. 3rd ACM Conference on Learning at
  Scale}, 2016, pp. 265--268.

\bibitem{cobos2020proposal}
R.~Cobos and J.~Soberón, ``{A Proposal for Monitoring the Intervention
  Strategy on the Learning of MOOC Learners},'' in \emph{Proc. CEUR Workshop},
  2020, p. 61–72.

\bibitem{cobos2021improving}
R.~Cobos and J.~C. Ruiz-Garcia, ``{Improving Learner Engagement in MOOCs using
  a Learning Intervention System: A Research Study in Engineering Education},''
  \emph{Computer Applications in Engineering Education}, vol.~29, no.~4, pp.
  733--749, 2021.

\bibitem{pascual2022proposal}
\BIBentryALTinterwordspacing
I.~Pascual and R.~Cobos, ``A proposal for predicting and intervening on mooc
  learners' performance in real time,'' in \emph{LASI-Spain 2022}, vol.
  Vol-3238, 2022. [Online]. Available:
  \url{http://ceur-ws.org/Vol-3238/paper4.pdf}
\BIBentrySTDinterwordspacing

\bibitem{karademir2022designing}
O.~Karademir, A.~Ahmad, J.~Schneider, D.~Di~Mitri, I.~Jivet, and H.~Drachsler,
  ``Designing the learning analytics cockpit - a dashboard that enables
  interventions,'' in \emph{Methodologies and Intelligent Systems for
  Technology Enhanced Learning, LNNS}, vol. 326.\hskip 1em plus 0.5em minus
  0.4em\relax Springer, 2022.

\bibitem{verbert2014learning}
K.~Verbert, S.~Govaerts, E.~Duval, J.~L. Santos, F.~Van~Assche, G.~Parra, and
  J.~Klerkx, ``Learning dashboards: an overview and future research
  opportunities,'' \emph{Personal and Ubiquitous Computing}, vol.~18, pp.
  1499--1514, 2014.

\bibitem{matcha2019systematic}
W.~Matcha, D.~Ga{\v{s}}evi{\'c}, A.~Pardo \emph{et~al.}, ``A systematic review
  of empirical studies on learning analytics dashboards: A self-regulated
  learning perspective,'' \emph{IEEE Transactions on Learning Technologies},
  vol.~13, no.~2, pp. 226--245, 2019.

\bibitem{hernandez2019edbb}
J.~Hernandez-Ortega, R.~Daza, A.~Morales \emph{et~al.}, ``{edBB: Biometrics and
  Behavior for Assessing Remote Education},'' in \emph{Proc. AAAI Workhop on
  Artificial Intelligence for Education}, 2020.

\bibitem{baro2018integration}
X.~Bar{\'o}-Sol{\'e}, Guerrero-Roldan \emph{et~al.}, ``Integration of an
  adaptive trust-based e-assessment system into virtual learning
  environments—the tesla project experience,'' \emph{Internet Technology
  Letters}, p. e56, 2018.

\bibitem{morales2016kboc}
A.~Morales, J.~Fierrez, R.~Tolosana, J.~Ortega-Garcia, J.~Galbally,
  M.~Gomez-Barrero, A.~Anjos, and S.~Marcel, ``Keystroke biometrics ongoing
  competition,'' \emph{IEEE Access}, vol.~4, pp. 7736--7746, 2016.

\bibitem{sharma2020eye}
K.~Sharma, M.~Giannakos, and P.~Dillenbourg, ``Eye-tracking and artificial
  intelligence to enhance motivation and learning,'' \emph{Smart Learning
  Environments}, vol.~7, no.~1, pp. 1--19, 2020.

\bibitem{ha2015tracking}
K.~Ha, I.-H. Jo, S.~Lim, and Y.~Park, ``Tracking students’ eye-movements on
  visual dashboard presenting their online learning behavior patterns,'' in
  \emph{Emerging Issues in Smart Learning}.\hskip 1em plus 0.5em minus
  0.4em\relax Springer, 2015, pp. 371--376.

\bibitem{2018_INFFUS_MCSreview1_Fierrez}
J.~Fierrez, A.~Morales, R.~Vera-Rodriguez, and D.~Camacho, ``Multiple
  classifiers in biometrics. {P}art 1: Fundamentals and review,''
  \emph{Information Fusion}, vol.~44, pp. 57--64, November 2018.

\bibitem{giannakos2019multimodal}
M.~N. Giannakos, K.~Sharma, I.~O. Pappas, V.~Kostakos, and E.~Velloso,
  ``Multimodal data as a means to understand the learning experience,''
  \emph{Intl. J. of Information Management}, vol.~48, pp. 108--119, 2019.

\bibitem{spikol2018supervised}
D.~Spikol, E.~Ruffaldi, G.~Dabisias, and M.~Cukurova, ``Supervised machine
  learning in multimodal learning analytics for estimating success in
  project-based learning,'' \emph{Journal of Computer Assisted Learning},
  vol.~34, no.~4, pp. 366--377, 2018.

\bibitem{Azevedo2022}
R.~Azevedo \emph{et~al.}, ``Lessons learned and future directions of metatutor:
  Leveraging multichannel data to scaffold self-regulated learning with an
  intelligent tutoring system,'' \emph{Frontiers in Psychology}, vol.~13, 2022.

\bibitem{ouyang2023artificial}
F.~Ouyang, W.~Xu, and M.~Cukurova, ``An artificial intelligence-driven learning
  analytics method to examine the collaborative problem-solving process from
  the complex adaptive systems perspective,'' \emph{International Journal of
  Computer-Supported Collaborative Learning}, pp. 1--28, 2023.

\bibitem{cukurova2020modelling}
M.~Cukurova, Q.~Zhou, D.~Spikol, and L.~Landolfi, ``Modelling collaborative
  problem-solving competence with transparent learning analytics: is video data
  enough?'' in \emph{Proceedings of the tenth international conference on
  learning analytics \& knowledge}.\hskip 1em plus 0.5em minus 0.4em\relax ACM,
  2020, pp. 270--275.

\bibitem{daza2023edbb}
R.~Daza, A.~Morales, R.~Tolosana, L.~F. Gomez, J.~Fierrez, and
  J.~Ortega-Garcia, ``{edBB-Demo: Biometrics and Behavior Analysis for Online
  Educational Platforms},'' in \emph{Proc. AAAI Conference on Artificial
  Intelligence (Demonstration)}, 2023.

\bibitem{daza2022alebk}
R.~Daza, D.~DeAlcala, A.~Morales, R.~Tolosana, R.~Cobos, and J.~Fierrez,
  ``{ALEBk: Feasibility Study of Attention Level Estimation Via Blink Detection
  Applied to e-learning},'' in \emph{Proc. AAAI Workshop on Artificial
  Intelligence for Education}, 2022.

\bibitem{daza2023matt}
R.~Daza, L.~F. Gomez, A.~Morales, J.~Fierrez, R.~Tolosana, R.~Cobos, and
  J.~Ortega-Garcia, ``{MATT: Multimodal Attention Level Estimation for
  e-learning Platforms},'' in \emph{Proc. AAAI Workshop on Artificial
  Intelligence for Education}, 2023.

\bibitem{daza2020mebal}
R.~Daza, A.~Morales, J.~Fierrez, and R.~Tolosana, ``{mEBAL: A Multimodal
  Database for Eye Blink Detection and Attention Level Estimation},'' in
  \emph{Proc. Intl. Conf. on Multimodal Interaction}, 2020, pp. 32--36.

\bibitem{hernandez2020heart}
J.~Hernandez-Ortega, R.~Daza, A.~Morales, J.~Fierrez, and R.~Tolosana, ``{Heart
  Rate Estimation from Face Videos for Student Assessment: Experiments on
  edBB},'' in \emph{Proc. Annual Computers, Software, and Applications
  Conference (COMPSAC)}, 2020, pp. 172--177.

\bibitem{2020_CDS_HCIsmart_Acien}
A.~Acien, A.~Morales, R.~Vera-Rodriguez, J.~Fierrez, and O.~Delgado,
  ``Smartphone sensors for modeling human-computer interaction: General outlook
  and research datasets for user authentication,'' in \emph{IEEE Conf. on
  Computers, Software, and Applications (COMPSAC)}, July 2020.

\bibitem{chong2020detecting}
E.~Chong, Y.~Wang, N.~Ruiz, and J.~M. Rehg, ``Detecting attended visual targets
  in video,'' in \emph{Proceedings of the IEEE/CVF Conference on Computer
  Vision and Pattern Recognition}, 2020, pp. 5396--5406.

\bibitem{2023_SNCS_multiAI}
A.~Peña, I.~Serna, A.~Morales, J.~Fierrez \emph{et~al.}, ``Human-centric
  multimodal machine learning: Recent advances and testbed on {AI}-based
  recruitment,'' \emph{SN Computer Science}, 2023.

\bibitem{2018_INFFUS_MCSreview2_Fierrez}
J.~Fierrez, A.~Morales, R.~Vera-Rodriguez, and D.~Camacho, ``Multiple
  classifiers in biometrics. {P}art 2: Trends and challenges,''
  \emph{Information Fusion}, vol.~44, pp. 103--112, November 2018.

\end{thebibliography}

\end{document}